# Security Awareness of End-Users of Mobile Health Applications: An Empirical Study

Bakheet Aljedaani[1], Aakash Ahmad[2], Mansooreh Zahedi[1], M. Ali Babar[1]
[1]The University of Adelaide, Adelaide, SA 5005, Australia
[2]University of Ha'il, Ha'il 2440, Saudi Arabia
[1][bakheet.aljedaani | mansooreh.zahedi | ali.babar]@adelaide.edu.au, [2]a.abbasi@uoh.edu.sa

## ABSTRACT
Mobile systems offer portable and interactive computing – empowering users – to exploit a multitude of context-sensitive services, including mobile healthcare. Mobile health applications (i.e., mHealth apps) are revolutionizing the healthcare sector by enabling stakeholders to produce and consume healthcare services. A widespread adoption of mHealth technologies and rapid increase in mHealth apps entail a critical challenge, i.e., lack of security awareness by end-users regarding health-critical data. This paper presents an empirical study aimed at exploring the security awareness of end-users of mHealth apps. We collaborated with two mHealth providers in Saudi Arabia to gather data from 101 end-users. The results reveal that despite having the required knowledge, end-users lack appropriate behaviour , i.e., reluctance or lack of understanding to adopt security practices – compromising health-critical data with social, legal, and financial consequences. The results emphasize that mHealth providers should ensure *security training* of end-users (e.g., threat analysis workshops), promote *best practices* to enforce security (e.g., multi-step authentication), and *adopt suitable mHealth apps* (e.g., trade-offs for security vs usability). The study provides empirical evidence and a set of guidelines about security awareness of mHealth apps.

## 1. INTRODUCTION
Mobile systems support portable computing that enable users to exploit context-sensitive services such as social networking, crowd-sensing, mobile commerce, and healthcare [1]. Mobile healthcare (i.e., mHealth) relies on (i) embedded sensors of a device (hardware to sense health-critical data), (ii) mobile apps (software to manipulate sensed data), and (iii) networking technologies (protocols to wirelessly transmit data). World Health Organization (WHO) refers to mHealth as medical healthcare practices, enabled via pervasive technologies, to facilitate stakeholders (e.g., health units, medics, patients) that provide or utilize healthcare services in an automated, efficient, and reliable manner. A widespread adoption of mobile computing has resulted in a rapid proliferation of mHealth apps that range from general apps such as decision support, vitals, and reproductive health to fitness monitoring apps for activity tracking and nutrition management [2]. The usage of mHealth apps by healthcare practitioners and patients is on a rise with 350,000 such apps available in two of the major app repositories provided by Android and iOS platforms [3]. Research2Guidance (R2G), a consultancy organization for mHealth technologies, reports that 78,000 new mHealth apps were added to apps stores in 2017 [4] with market revenue for digital health expected to reach USD 31 billion by 2020 [5]. Despite the offered benefits [6] and expected revenues [5] of mHealth systems, security of health-critical data remains a challenge for sustained growth and mass-scale adoption of mHealth apps [7-10]. The interest of attackers in health-critical data (pulse rate, blood-pressure, disease symptoms, etc.) has increased due to its value in the 'black market' as well as social, legal, and financial consequences of compromised data [11]. According to the Ponemon Institute, the price per single medical record escalated from $369 in 2016 to $380 in 2017 due to the regulations and their implementations to secure health-critical data [12].

Security of mHealth apps is a critical challenge due to pervasive environment in which mobile devices continuously ingest health-critical data from embedded sensors, process and persist data inside the device, and transmit it across ad-hoc networks [8, 13]. Experts believe that technical solutions such as authentication and multi-step authorization cannot address security issues alone, instead the role of end-users and their understanding of security related issues is essential to ensure secure mobile computing [14]. Some recent studies have highlighted that social engineering method can be used by hackers to deceive end-users into leaking their private information [15, 16]. Such security lapses are common in mHealth solutions as the end-users usually lack sufficient awareness about security configurations, critical warnings, and consequences of security breaches [17]. A recently conducted study of secure mHealth app development has highlighted that developers who follow a predefined software development lifecycle (SDLC) often assume that they have already delivered a secure app [18]. However, end-users may find security features hard to understand, get deceived by hackers, or misled by app permissions to disclose private and sensitive information [16]. Due to different categories of mHealth apps (e.g., clinical health, fitness monitoring, and health diagnostics) end-users behave differently depending of the type of apps being used and their security awareness [19, 20]. The existing research on this topic, such as [7, 8, 10], indicate that security of mHealth apps lags behind the capabilities of hackers that target mHealth apps. In a recent study [18], the authors present developers' perspectives on the challenges, recommended practices, and motivators for developing secure mHealth apps. To date, there has been no empirical effort aimed at investigating the security awareness of end-users towards the usage of mHealth apps for clinical settings. The study presented in this paper complements the existing research on developers' perspective [18] by empirically investigating end-users' security awareness of mHealth apps that contain health-critical and other personal data.

We conducted an empirical study - collaborating with two commercial mHealth providers[1] in Saudi Arabia - to survey 101 end-users of mHealth apps for investigating the level of security awareness among end-users. Security awareness of end-users refers to the *'required human knowledge, attitude, and behaviour to understand the potential security risks, their implications, and available countermeasures for securing mHealth'* [15]. The surveyed users are affiliated in different professional roles such as clinical practitioners, medical doctors, nurses, and healthcare supervisors with the above mentioned health providers. Demography analysis of the end-users highlight their diverse educational backgrounds, IT skills, years of experience with usage of various mHealth apps on different mobile devices compatible with major mobile computing platforms including Android and iOS. For this study, we utilized the Human Aspects of Information Security Questionnaire (HAIS-Q) [17] that allows to measure security awareness through Knowledge, Attitude and Behaviour (KAB) model [21]. To measure and understand the security-awareness, we formulated the following Research Questions (RQs):

*RQ1: What is the level of security knowledge, attitude, and behaviour of end-users about mHealth apps?*

*RQ2: What are the relationships between security knowledge, attitude and behaviour? And how do they influence the end-users while utilizing mHealth apps?*

To analyze the collected data for answering the RQs, we used statistical methods, including descriptive analysis, data correlation of survey statements, and regression testing of the recorded responses. The results highlight that end-users' security knowledge strongly influences their attitude (i.e., they understand *what potential threats are*). However, end-users' security knowledge has no significant impact on their behaviour (i.e., they lack necessary actions about *how to protect data from potential threats*). Security training, recommended practices, and adoption of appropriate mHealth apps, both at individual and organization level are the key to secure utilization of mHealth services.

The study provides empirical evidence about the level of end-users' security awareness, i.e., (i) *knowledge* about recurring security threats, (ii) *attitude* to securing their data, and (iii) *behaviour* to mitigate the threats by adopting security practices. This study provides empirical evidence and a set of guidelines to facilitate researchers, practitioners, and stakeholders to develop and adopt secure mHealth apps for clinical practices and public health.

## 2. RELATED WORK

We now review the related work generally classified as (i) security awareness of end-users (Section 2.1), and (ii) approaches to measure security awareness (Section 2.2). The review helps us to motivate this study and introduces terminologies and concepts that are used in this study.

## 2.1 Security Awareness of End-Users of mHealth Apps

mHealth apps to support clinical practices are primarily for commercial purposes to enrich users' experience by offering customized functionality and automating trivial tasks of healthcare [13]. Currently, healthcare providers are adopting such mHealth apps to empower patients by providing healthcare services that are pervasive, readily available, efficient, and cost-effective [13, 16]. For example, many healthcare services such as diagnosing cardio rate, measuring blood pressure, or detecting fever can be performed at users' end to eliminate prior appointments or personal visits to health units. Moreover, mHealth apps enable end-users to electronically view, share, and manage their medical history, including but not limited to lab reports, radiology images, and scheduled appointments [22]. Despite the offered benefits, patients' health-critical data is at risk either due to security flaws in an app [7, 8, 10] or lack of security awareness by app users [7, 16].

*Lack of security awareness:* A study reported in [19] engaged 24 focus groups with more than 250 participants to examine end-users' attitudes and perceptions regarding mHealth systems being used at healthcare centers. The study revealed that end-users' attitudes were highly contextualized towards security depending on the type of information being communicated, rational for such communication, and consumers of shared information. From the app development perspective, the security of mHealth apps can be enhanced by implementing different control mechanisms (e.g., encryption, authentication, secure storage, access control) that support confidentiality and integrity of data [23]. The adoption of mHealth apps by health providers as well as end-users is on a steady rise, however; some recent studies have reported that low knowledge of security features for end-users is still an issue [16, 24-26]. The authors in [27] pointed out that today's smartphones implement many security features that range from device lock mechanisms to remote data wiping or end to end encryption. However, even the most advanced systems or sophisticated features of security cannot guarantee users' behaviour and actions that enhance or compromise security and privacy of their classified data [19, 28]. A recent study of more than 450 smartphone owners indicates that they do not use or are unaware of security features provided by default [28]. In the case of mHealth systems, patients as end-users might grant more permissions than necessary to unintentionally share their health data or allow other apps to unnecessarily access it [16]. A recently conducted mapping study [29] also indicates that, despite implementing the state-of-the-art security features for mHealth app, lack of knowledge about the security features or privacy permissions by end-users can compromise personal information and health-critical data. The mapping study reviews 365 studies (published research) and suggests that security and privacy specific education and training of app developers and users are two of the most critical factors to support secure development and usage of mHealth systems.

*Enabling security awareness:* Developers, providers, and consumers (commercial enterprises) need to play their roles in increasing the end-users' security awareness [16, 26]. For examples, commercial enterprises as consumers of mHealth systems can provide training and guidelines to explain security features of an app to it's end-users (e.g., doctors, nurses, clinical technicians, patients). Similarly, developers of mHealth apps can engineer their apps to support effective security decisions and facilitate (or enforce) end-users to follow security practices. A typical example of security enforcement by an app can be password management schemes that refuse to accept a weak password or

---

multi-step authentication to access private data [30]. To enhance security awareness, mHealth apps can explicitly indicate access permissions that are essential or optional for end-user to select. As opposed to the adaptive security [31] that enables apps and devices to configure their security protocols at runtime, human-centric security focuses on secure development and usage practices (i.e., users' actions) to enhance the security of mobile systems [17, 18].

## 2.2 Approaches to Measure End-Users' Security Awareness

Security awareness is mostly measured by end-users feedback via survey or questionnaire-based studies such as [17, 32]. The authors of [26] conducted simulation exercises as controlled experiments to understand participants' behaviours and reaction corresponding to their security awareness about potential threats. Survey or simulation-based approaches to measure security awareness can be helpful to identify users' perspective and behaviours, highlighting human intellect, attitude, and weaknesses while facing potential security threats [22, 23, 26].

*Models for measuring security-awareness:* HAIS-Q is a well-known approach that has been used and validated in several studies [17, 21] to measure information security awareness for a diverse group of users (e.g., students, professionals, and volunteers). Some studies have followed the HAIS-Q guidelines [14] to measure security awareness of smartphone users [33]. The Knowledge, Attitude, and Behaviour (KAB) model that underpins HAIS-Q aims to investigate psychological aspects and their impact on (information security) awareness [34]. KAB model refers to a representative study of a specific population that collects information about what is known, believed, and done about a particular topic [35]. Knowledge refers to the level of security knowledge that users already know, an attitude refers to how the users feel about the knowledge, and behaviour refers to the actions that users may perform to ensure security [17].

The existing studies, e.g. [19, 20], focus on users' security perceptions of fitness monitoring apps such as activity tracking, step counters, and cardio rate analysis. Unlike the fitness monitoring apps, clinical mHealth apps such as patient management systems handle highly sensitive health-critical data and personal information. The scope of this study is to investigate security awareness and guidelines for the class of clinical health systems in the context of mHealth. From an operational perspective, clinical health apps collect patients' health data, personal information, and medical history that is electronically shared across health care units among different medical professionals – involving a multitude of security issues. To the best of our knowledge, there is no empirical study to investigate end-users' perspectives and security awareness toward using clinical mHealth apps.

## 3. RESEARCH METHOD

In this section, we discuss the research method that comprises of four phases to conduct this study. The overall research method and its individual phases are illustrated in Figure 1.

### 3.1 Phase I – Design the Study Protocol

We developed the study's protocol in the initial phase, shown in Figure 1, that comprises of three steps including (i) specification of research questions (ii) designing survey questionnaire, and (iii) identify data access methods. As part of this phase, we performed the literature review (Section 2) to devise two RQs (Section 1). The RQs drove the design of the survey questionnaire, which was based on the KAB model [33] that underpins HAIS-Q [17, 21]. The literature review also pointed out that human aspects, i.e., end-users are concerned about four issues (a) *what* methods are used to access their health-critical data, (b) *who* can access their data, for (c) *what* purposes their data is being accessed, and (d) *how* their data is being stored and transmitted. Figure 2 shows four security mechanisms (i.e., solutions) that can be classified as eight security critical scenarios (i.e., instance of a solution). For each security critical scenario, we formulated a security knowledge statement, an attitude statement and a behaviour statement as in Figure 2. For example, as per Figure 2, a knowledge statement can be expressed as *'I should use strong password (not easy to guess) for mHealth app'*. A total of 23 statements (S1 – S23) were formulated that are classified as 08 statements for measuring security knowledge, 08 statements to measure end-users' attitude, and 07 statements for measuring end-users' behaviour toward using mHealth apps. We provided three options in our survey questionnaire (i.e., True, False, and Don't know) to the end-users. It is worth mentioning that we excluded statement about data encryption from the behaviour since encrypting data is an automated process of an app and does not represent an actionable task for end-users. Figure 2 provides an example of how we measured the security awareness of mHealth apps end-users using the KAB model [33].

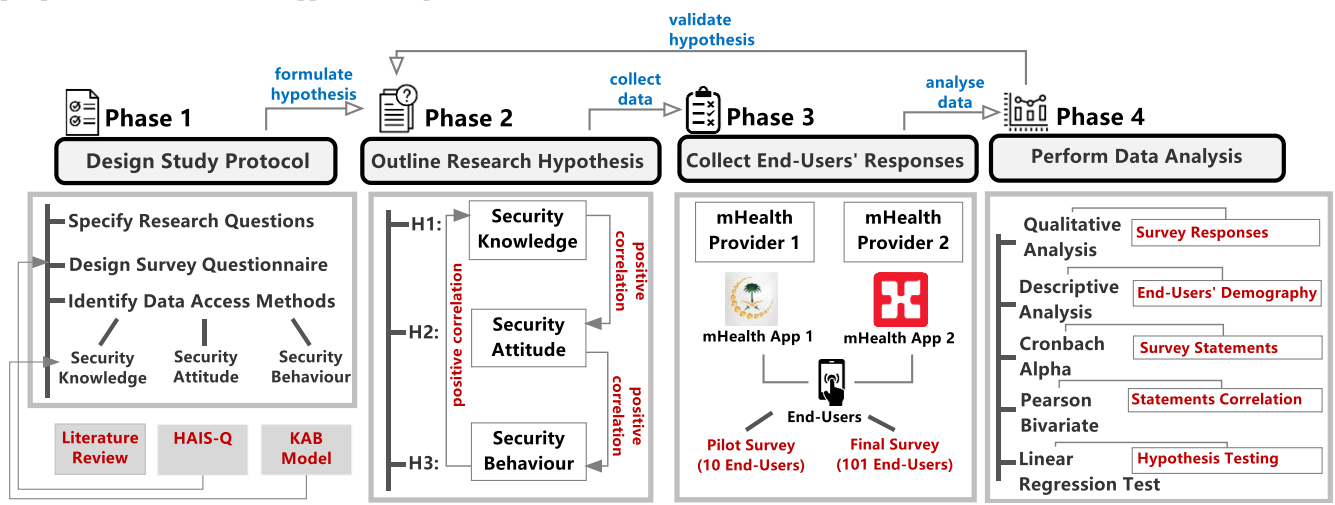

**Figure 1 - Overview of the Research Methodology**

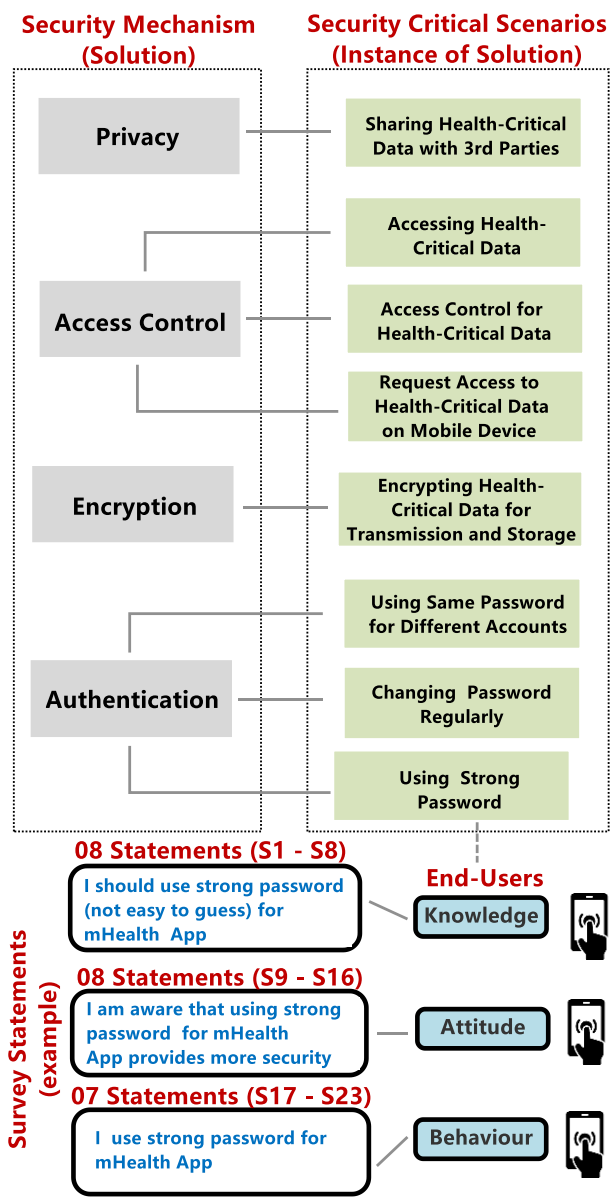

**Figure 2 - Overview of Security Awareness Measurement for End-Users using the KAB Model.**

## 3.2 Phase II – Outline Research Hypothesis

We outlined three research hypothesis and correlation between them to validate the survey findings as in Figure 1 (Phase 2). The research in [21] confirms that raising the knowledge level of information security and procedures would increase end-users' attitude towards information security policy and procedures, which should translate into more risk-averse information security behaviour. Based on the guidelines from [30], we outlined the following hypotheses (H1 to H3):

*H1: Security knowledge of the end-users of mHealth apps has a positive correlation with their attitude.*
*H2: Attitude of the end-users of mHealth apps is positively correlated with their behaviour.*
*H3: Security knowledge for the end-users of mHealth apps has a positive correlation with their behaviour.*

The hypotheses positively correlate end-users' security knowledge, attitude, and behaviour to measure their security awareness for mHealth apps. The hypotheses can be tested based on qualitative analysis of the survey data from end-users of mHealth apps that is detailed next.

## 3.3 Phase III – Collect End-Users' Responses

The third phase of methodology is focused on data collection from end-users of two major mHealth providers. End-users' data via survey questionnaire was collected in collaboration with two mHealth providers namely (i) KFMC - *King Fahad Medical City* (having **iKFMC app**) and (ii) HMG - *Dr. Sulaiman Al Habib Medical Group* (having **Dr. Sulaiman Alhabib app**). The link to the survey-based questionnaire, responses from end-users, and extended details demographic information are provided in [36]. These health providers were purposefully chosen because they provide mHealth apps to allow their users to access a wide range of services such as creating, storing, and sharing medical records, viewing scanned images, lab results, and automating their clinical practices. The first author personally visited both mHealth providers during January and February 2020 to carry out the study and collect data. All of the end-users that we surveyed and both the mHealth providers were based in Saudi Arabia thus limiting the geo-distribution of survey participants and its impact on study findings. Some travel restrictions globally, during the said time-period, also limited our planned on-site data collection from more countries. Extended work is in progress - designing and deploying web-based survey and a case study - to seek feedback from geographically diverse end-users and their healthcare providers. Potential respondents were asked to join our study while they were waiting for their turns to see their doctors. A brief explanation for our research was given, and we assured them that they are free to withdraw anytime, should they like to. We ensured that all our respondents were: i) $\geq$ 18-year-old, and ii) currently using the provided mHealth apps. We conducted pilot testing for our questionnaire with ten respondents to mitigate any of their concerns (e.g., wording problems, statement ambiguity). After removing incomplete responses, we obtained 101 responses from the end-users of KFMC and HMG. At the beginning of our survey, we asked our respondents to answer a few questions related to their background (demographic details), followed by the main topics under investigation (security awareness). Figure 3 (a - f) presents a summary of our respondents' demographic information including (1) *mobile platforms* and (2) *mobile devices* of end-users, their (3) *gender classification*, (4) *age group*, (5) *IT knowledge level*, (6) *formal education*, and (7) *mHealth app usage*. The participants' demographic information is used to contextualize the responses to the main questions of the survey. For example, analyzing the age, level of IT knowledge, or education level (e.g., Bachelors' degree, Diploma) of end-users' can help better understand security-awareness for a specific group of users.

## 3.4 Phase IV – Perform Data Analysis

We used a well-known data analysis software, SPSS version 27 (IBM), for the quantitative analysis of our collected data. Descriptive analysis, along with mean and Standard Deviation (SD) were conducted to report the respondents' demographic data (Figure 3) and the survey responses (Phase III). We calculated the Cronbach Alpha to measure the reliability and internal consistency of our survey statements. To test the research hypotheses (Phase III), we performed two linear regression tests, firstly, to predict the respondents' attitude according to their mean knowledge, and secondly, to predict the respondents' behaviour based on their mean knowledge and mean attitude.

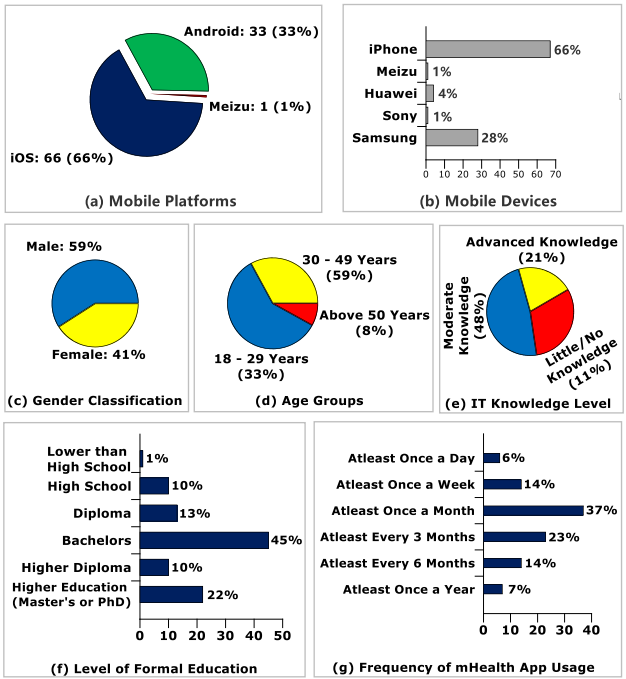

**Figure 3 - Demographic Details of mHealth Apps End-Users (Sample Size =101)**

*Ethics Approvals:* Both health providers granted us permission to use end-users' feedback and other relevant data, for research purposes only, through their Institutional Review Board (KFMC approval number: **19-462E**, HMG approval number: **HAP-01-R-082**). Furthermore, our study is approved by the Human Research Ethics Committee at the University of Adelaide (approval number **H-2019-165**). Further details of the methodology, statistical data analysis and ethics approval are in [36].

## 4. RESULTS OF END-USERS' SECURITY AWARENESS

We now present the results that reflect end-users' security awareness towards the usage of mHealth apps. The results provide answers to RQ-1s in Section 1, and RQ-2 in Section 4.2.

### 4.1 End-Users' Security Knowledge, Attitude, and Behaviour (RQ-1)

As per the research methodology (Phase 4, Fig. 1), we applied a number of statistical methods to formalize data analysis (measuring consistency, correlation, and description) of end-users' responses [36] presented in Table 1 – Table 5.

*A. Measuring Consistency of Survey Statements:* To answer RQ1, Table 1 below presents Cronbach's alpha to measure the internal consistency of the survey statements, i.e., how our survey statements are closely related to measure end-user security knowledge, attitude, and behaviour while using mHealth apps [37]. Cronbach's alpha helped us to determine the reliability of the measurement values (end-users' responses that are captured using multiple Likert questions). Table 1 highlights that the 3 constructs each having a specific value (i) security knowledge = 0.820, (ii) attitude (towards security) = 0.732, and (iii) behaviour (while using mHealth app) = 0.722. The values of Cronbach's alpha were obtained by calculating the results of each dimension (e.g., only the attitude related to the statements were calculated together). Our results indicate that we met the minimum acceptable coefficient value (i.e., $\geq 0.7$).

**Table 1. Cronbach's alpha for end-users' Knowledge, Attitude, and Behaviour.**

| Constructs | Cronbach's alpha |
|---|---|
| Knowledge | 0.820 |
| Attitude | 0.732 |
| Behaviour | 0.722 |

*B. Measuring Correlation between Survey Statements:* We conducted Pearson product-moment correlation (Pearson's correlation, for short) [38] to test the correlation (i.e., strength and direction of association) between the survey statements for obtained data, presented in Table 2 – Table 4. Specifically, Table 2 – Table 4 show the correlation strength and significance among the survey statements for each dimension of security awareness (i.e., knowledge, attitude, and behaviour). All our survey statements showed a positive relationship ranging from strong (i.e., 0.86: *access to health data*) to a very weak (i.e., 0.02: *changing the password regularly*) relationship, as in Table 2. Only two statements in the attitude dimension (i.e., *encrypting health data during transmission and storage*, and *changing the password regularly*) showed a very weak negative relationship, as in Table 3. On the other side, our test results imply that 61% of our survey statements had a significant relationship (i.e., 48% at the 0.01 level, and 13% at the 0.05 level). This means that about 39% of our survey statements showed that there was no significant relationship between end-users security knowledge, attitude, and behaviour. For those reasons which are related to the correlation strength and their significance, we ensured that our survey items are measuring the interest through reviewing and clarifying any ambiguity.

*C. Descriptive Analysis of Responses:* Table 5 presents a descriptive summary of end-users' responses (**R1** – **R101**) to the given (**S1** – **S23**) in our survey [36]. For quantification and simplification of data analysis, we assigned numbers to the given options (i.e., True=1, Don't know=2, False=3). Frequency, percentage, mean and Standard Deviation (SD) are presented in Table 5. Since the middle value is 2, a mean of less than 2 implies that the respondents have agreed to the statement (True = 1). Whereas, a mean of greater than 2 implies that respondents have disagreed with the statement (False = 3). Our findings indicate that end-users have different opinions about (i) *what* they need to know (i.e., knowledge), (ii) *why* they should know (i.e., attitude) and (iii) *how* they should behave (i.e., behaviour) when it comes to using mHealth apps. For example, end-to-end data encryption (e.g., using Transport Layer Security (TLS) with 128-bit encryption) is a method to secure data that travels across devices, over various networks, and accessed by a third-party. We provided two statements (S5 and S13) to examine our respondents' knowledge and attitude toward applying encryption within the used mHealth apps. Only 43/101 (i.e., 42.6%) of end-users believed that they should know whether or not their health data, which have been collected by mHealth apps, is sent and stored in an encrypted format. Whereas, 47/101 (i.e., 46.5%) of our respondents indicated that they are not aware whether or not their health data, which have been collected by mHealth apps, are encrypted during transmission and storage. In fact, one respondent (**R33**) commented that *"[...] the responsibility to protect data is one of the patient's rights to be fulfilled by health service provider"*. Thus, there is a need at app providers' end to share information with end-users about the existing features which make the apps more secure and trustable.

Our results indicate that the respondents' level of security knowledge, attitude, and behaviour towards using mHealth app

**Table 2. Correlation for Security Knowledge Statements**

| Security Implementation Strategies | 1 | 2 | 3 | 4 | 5 | 6 | 7 | 8 |
|---|---|---|---|---|---|---|---|---|
| 1. Sharing health data with third parties. | 1 | | | | | | | |
| 2. Accessing health data. | .86** | 1 | | | | | | |
| 3. Controlling access to health data. | .64** | .64** | 1 | | | | | |
| 4. Requesting access to the data on the phone. | .67** | .56** | .44** | 1 | | | | |
| 5. Encrypting health data during transmission and storage. | .47** | .54** | .35** | .50** | 1 | | | |
| 6. Using the same password for different accounts. | .27** | .31** | .18 | .16 | .16 | 1 | | |
| 7. Using a strong password. | .27** | .33** | .35** | .20* | .17 | .71** | 1 | |
| 8. Changing the password regularly. | .24** | .19 | .17 | .31** | .02 | .15 | .33** | 1 |

**Table 3. Correlation for Attitude Statements**

| Security Implementation Strategies | 1 | 2 | 3 | 4 | 5 | 6 | 7 | 8 |
|---|---|---|---|---|---|---|---|---|
| 1. Sharing health data with third parties. | 1 | | | | | | | |
| 2. Accessing health data. | .57** | 1 | | | | | | |
| 3. Controlling access to health data. | .53** | .53** | 1 | | | | | |
| 4. Requesting access to the data on the phone. | .51** | .37** | .45** | 1 | | | | |
| 5. Encrypting health data during transmission and storage. | .35** | .24* | .21* | .32** | 1 | | | |
| 6. Using the same password for different accounts. | .18 | .05 | .25* | .22* | .18 | 1 | | |
| 7. Using a strong password. | .08 | .17 | .25* | .02 | .06 | .42** | 1 | |
| 8. Changing the password regularly. | .04 | .19 | .11 | .21* | -.05 | .14 | .42** | 1 |

**Table 4. Correlation for Behaviour Statements**

| Security Implementation Strategies | 1 | 2 | 3 | 4 | 5 | 6 | 7 |
|---|---|---|---|---|---|---|---|
| 1. Sharing health data with third parties. | 1 | | | | | | |
| 2. Accessing health data. | .66** | 1 | | | | | |
| 3. Controlling access to health data. | .52** | .58** | 1 | | | | |
| 4. Requesting access to the data on the phone. | .48** | .42** | .49** | 1 | | | |
| 5. Using the same password for different accounts. | .18* | .22* | .09 | .06 | 1 | | |
| 6. Using a strong password. | .15 | .18 | .17 | .17 | .41** | 1 | |
| 7. Changing the password regularly. | .13 | .20* | .04 | .18 | .18 | .26** | 1 |

**. Correlation is significant at the 0.01 level (2-tailed). *. Correlation is significant at the 0.05 level (2-tailed).

vary. Such variations are primarily due to a number of factors that are presented in demographic details of the end-users (Figure 3). This indicates that respondents were not confident about some of the existing security measures implemented in the mHealth apps. The results indicate security specific documentation or education that must be provided to end-users from mHealth app providers.

## 4.2 Correlation between End-Users Security Knowledge, Attitude, and Behaviour (RQ-2)

To answer RQ-2, i.e., measuring the correlation, we utilized the KAB model (Figure 2) that measures end-user security awareness by examining three dimensions to test the outlined hypotheses (H1 – H3). Therefore, we tested the hypothesis that there is a significant positive correlation between (a) the respondents' knowledge which affects their attitude, and (b) the respondents' knowledge and attitude, which affect their behaviour. We assigned the predictor variables to give us an estimation of the significance, and for that, we conducted regression analysis using SPSS software. As per the KAB model [17, 21] to measure the correlation, we considered knowledge as an exogenous variable and considered attitude and behaviour as two endogenous variables [21]. Thus, we performed two linear regression tests (i.e., one simple regression and one multiple regression), both detailed below, to assess the correlation of our variables.

*A. Simple Regression Test - Predicting Attitude based on Knowledge:* A simple regression test is aimed to see whether we could predict attitude based on the knowledge that end-users had or not. Figure 4 presents our findings based on the correlation among knowledge, attitude and behaviour. Our analysis revealed that the knowledge variable had a statistically significant impact on the attitude variable producing $R^2_1$= 0.602, (B=0.718, t= 12.236, p < .001). This means that our respondents' knowledge accounted for 60.2% of the variance in their attitude, which implies that, respondents' knowledge is positively related to their attitude. For example, the survey statements (S14: *'I am aware that using one password for different accounts and apps will make it easy for me; but, it is insecure'*) was endorsed (True = 1) by 70% of the end-users. Similarly, 70% of the end-users agreed (S6: *'I should use different passwords for different accounts and apps'*). This correlation equated to B= 0.718 for H1: Security knowledge for the end-users of mHealth apps has a positive correlation with their attitude.

**Table 5. Descriptive Summary of Responses to the Statements in the Survey (Sample Size: N=101)**

| Statements | True (1), n (%) | Don't know (2), n (%) | False (3), n (%) | Mean (SD) |
|---|---|---|---|---|
| **Knowledge related statements** | | | | |
| S1: I should be informed when my health data, which have been collected by mHealth apps, are being shared with third parties, such as other hospitals or clinics. | 68 (67.3) | 17 (16.8) | 16 (15.8) | 1.49 (.756) |
| S2: I should know who accesses my health data and for what purpose. | 61 (60.4) | 25 (24.8) | 15 (14.9) | 1.54 (.742) |
| S3: I should have control of my health data. | 54 (53.5) | 25 (24.8) | 22 (21.8) | 1.68 (.812) |
| S4: I should know that some apps request access to my health data (e.g., marketing purposes). | 50 (49.5) | 35 (34.7) | 16 (15.8) | 1.66 (.739) |
| S5: I should know whether my health data, which have been collected by mHealth apps, is sent and stored in an encrypted format. | 43 (42.6) | 38 (37.6) | 20 (19.8) | 1.77 (.760) |
| S6: I should use different passwords for different accounts and apps. | 72 (71.3) | 16 (15.8) | 13 (12.9) | 1.42 (.711) |
| S7: I should use a strong password (not easy to guess) for the mHealth app. | 79 (78.2) | 14 (13.9) | 8 (7.9) | 1.30 (.609) |
| S8: I should change the password for the app regularly. | 60 (59.4) | 23 (22.8) | 18 (17.8) | 1.58 (.778) |
| **Attitude related statements** | | | | |
| S9: I am aware that my health data, which have been collected by mHealth apps, will not be shared with third parties, such as other hospitals or clinics. | 56 (55.4) | 23 (22.8) | 22 (21.8) | 1.66 (.816) |
| S10: I am aware of who accesses my health data and for what purpose. | 37 (36.6) | 36 (35.6) | 28 (27.7) | 1.91 (.801) |
| S11: I am aware that having control of my health data provides security. | 51 (50.5) | 25 (24.8) | 25 (24.8) | 1.74 (.833) |
| S12: I am aware that some apps request accessing health data more than they need. | 44 (43.6) | 37 (36.6) | 20 (19.8) | 1.76 (.764) |
| S13: I am aware that my health data, which have been collected by mHealth apps, are encrypted during transmission and storing. | 32 (31.7) | 47 (46.5) | 22 (21.8) | 1.90 (.728) |
| S14: I am aware that using one password for different accounts and apps will make it easy for me; but, it is insecure. | 71 (70.3) | 15 (14.9) | 15 (14.9) | 1.45 (.741) |
| S15: I am aware that using a strong password, that's not easy to guess, for mHealth app provides security. | 82 (81.2) | 8 (7.9) | 11 (10.9) | 1.30 (.656) |
| S16: I am aware that changing the password for the app regularly provides security. | 62 (61.4) | 17 (16.8) | 22 (21.8) | 1.60 (.826) |
| **Behaviour related statements** | | | | |
| S17: I get informed when my health data, which have been collected by mHealth apps, are being shared with third parties. | 49 (48.5) | 23 (22.8) | 29 (28.7) | 1.80 (.860) |
| S18: I know who access my health data and for what purpose. | 38 (37.6) | 30 (29.7) | 33 (32.7) | 1.95 (.841) |
| S19: I already have control of my health data. | 47 (46.5) | 24 (23.8) | 30 (29.7) | 1.83 (.861) |
| S20: I may accept or deny the request based on what I think is secure. | 45 (44.6) | 29 (28.7) | 27 (26.7) | 1.82 (.829) |
| S21: I use different passwords for different accounts and apps. | 64 (63.4) | 12 (11.9) | 25 (24.8) | 1.61 (.860) |
| S22: I use a strong password, that's not easy to guess for mHealth app. | 83 (82.2) | 5 (5.0) | 13 (12.9) | 1.31 (.689) |
| S23: I change my password of the app regularly. | 45 (44.6) | 9 (8.9) | 47 (46.5) | 2.02 (.959) |

*B. Multiple Regression Test - Predicting Behaviour based on Knowledge and Attitude:* The second regression tests the possibility of predicting end-users' behaviour based on knowledge and attitude. Our analysis also revealed that both knowledge and attitude have a statistically significant impact on behaviour producing $R^2_2 = 0.526$, as in Figure 4. The results of multiple regression test indicate that our respondents' knowledge and attitude account for 52.6% of the variance on their behaviour. Although, coefficient results indicate that attitude predictor is statistically significant, (B = .716, t = 5.713, p < .001), and yet, knowledge predictor is not statistically significant, (B = .125, t = 1.076, p < .285). For example, we notice that many respondents,

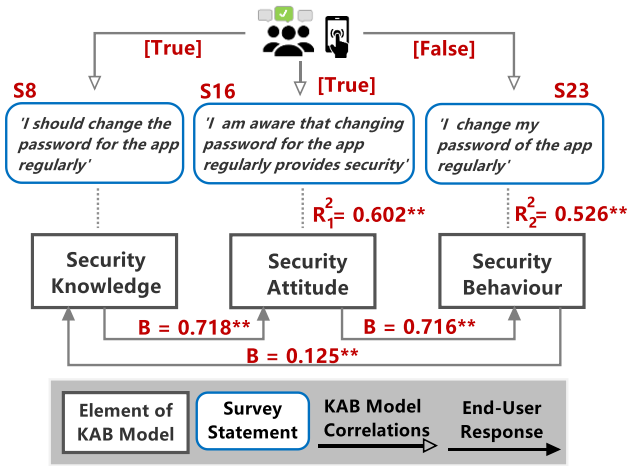

**Figure 4. Results in Support of the KAB Component of the HAIS-Q Model**

such as **R32**, **R34**, **R42**, **R45**, agreed that they knew and believe (i.e., knowledge and attitude) that mobile apps would be secure when using different passwords for different accounts and apps. However, their actual behaviour is not representing their knowledge and attitude. By reviewing the obtained data, particularly knowledge and behaviour responses, we notice a contradiction between what respondents should know (i.e., their knowledge) and how they should behave (i.e., their behaviour). For instance,

- *Knowledge and Attitude:* 59% of our respondents reported that they knew that changing the password for the app frequently (S8 in Table 5) would enhance the security.
- *Knowledge and Behaviour:* In reality, 47% of our respondents indicated that they did not change the password frequently, and thus, behaving according to the knowledge which they already had.

The results of the multiple regression test provide support (B = .716, p < .001) for H2: Attitude of the end-users of mHealth apps is positively correlated with their behaviour and reject (B = .125, p < .285) H3: Security knowledge for the end-users of mHealth apps has a positive correlation with their behaviour. Overall, our results can be more robust in case we could have involved more respondents (e.g., sample size N = 300). We believe that the outcomes support the used model, and it is acceptable to evaluate the end-users' security awareness toward using mHealth apps. The findings indicate that some of the respondents had good knowledge and attitude but did not demonstrate good behaviour, which could have exposed them to a significant risk. This is a clear indication of the need of providing end-users with suitable training to promote good security behaviour. Such strategy would also help to enhance their awareness when using mHealth apps.

## 5. DISCUSSION AND FUTURE WORK
We now discuss the key results of our study and highlight the potential future work to extend this study's findings.

### 5.1 Overall Security Awareness of End-Users
The answer to RQ-1 indicates end-users' security-awareness for mHealth apps varies depending on different factors such as educational backgrounds, prior experience, IT specific understanding and age groups (Figure 3). The analysis of the data in Table 3 and linking it with demographic information in Figure 3 indicates that end-users' of *age group* 30 – 49, who have *moderate or advanced knowledge of IT* systems and having at least a *bachelor's degree* are well aware of the security issues pertaining to their health-critical data and personal information. However, the security knowledge of end-users does not always translate to the required behaviour that can make their data more secure. For example, despite the awareness about security criticality of health data, end-users rarely choose complex passwords and feel it inconvenient to frequently change passwords. By combining demographic details and statistical data analysis, we imply that the level of security-awareness varies and it is primarily influenced by users' behaviour that can make their data more or less secure. RQ-1 aims to investigate human-centric view towards security-aware usage of mHealth apps with a focus on (i) what the end-users' know (i.e., knowledge), (ii) how they perceive security threats based on their knowledge (i.e., attitude), and (iii) why they should act to enable or enhance security (i.e., behaviour). This RQ does not answer what steps can be taken to enhance users' knowledge, change their attitudes, and motivate them to behave in a manner that ensures the security and privacy of health-critical data.

*Needs for future research:* It is interesting to observe that many respondents do not act according to their knowledge and attitude as shown in Table 2 – Table 3. This can be due to many factors such as the need of manual actions to enhance security (e.g., frequent password changes) and lack understanding for the consequences of security breaches. Future research can be focused on the effectiveness of formal training or security education, such as workshops, presentations, or hand on sessions that enhance security awareness of mHealth apps end-users. The scope of such research can go beyond mHealth apps to also investigate other mobile apps that deal with sensitive information (e.g., mobile banking apps). An investigation of the impact of security training on end-users' security awareness can help to identify best practices, processes, and patterns that can be adopted as guidelines. As part of the future research, we aim to explore the practical approaches for security and their impacts on end-users.

### 5.2 Correlation between KAB Model for Security Awareness
The study [21] has indicated that better knowledge is associated with better attitude, and both are associated with better behaviour, and that leads to better security awareness about security risks and counter-measures. The findings of RQ-2 indicate that our respondents' knowledge has a positive effect on their attitude, suggesting that the respondents have the necessary knowledge and attitude to ensure the security of mHealth apps. However, the analysis indicates that our respondents' knowledge and attitude does not affect their self-reported behaviour (Figure 4). The explanation for our findings can be related to organizational factors (e.g., lack of security policy and guidelines) and individual factors (e.g., personality). Since our study is more focused on the security *awareness of end-users toward using mHealth apps, the self-*reported behaviour results are slightly different than [21]. We targeted the respondents who are mainly using their own devices, and there are no assigned security policy and guidelines that they should follow. Unlike the scope of [21] that investigates how employees adhere to the existing security policy and guidelines when using organizations' computers for network-based applications.

*Needs for future research:* mHealth apps capture, process, and share health-critical data, and lack of security policy and guidelines for end-users is a critical challenge to be addressed. There is an urgent need of research efforts to develop appropriate security policies and guidelines for using mHealth apps so that end-users understand how to prevent security-related risks while using the apps. Furthermore, the security policy and guidelines would explain the right actions that end-users need to take in different circumstances. At the same time, providing suitable security awareness regarding the policy and guidelines for the end-users is just as important as developing secure mHealth apps.

## 6. THREATS TO VALIDITY
We now discuss some threats to the validity that highlight the certain assumptions and/or potential limitations of our study.

**Threat I –Data Collection about End-User' Behaviour:** Our study relies on survey-driven feedback (Phase 3, Figure 1) as self-reported knowledge, attitude, and behaviour of end-users that involves potential human bias in reporting. We tried to minimize this bias with the use of KAB [34] and HAIS-Q [17, 21] models, along with a pilot survey of 10/101 (approximately 10%) of the survey population. Based on the outlined hypothesis (Phase 2, Figure 1) and their validation (Section 4), respondents' knowledge had not influenced their behaviour towards mHealth security. We consider such correlation as an indication of fair responses by participants. In order to further minimize this threat, i.e., end-users' behaviour and self-reporting about mHealth security, there is a need to incorporate further validation. In order to achieve that, we propose to implement an app with some simulated attacks such as excessive permissions from app and notifications for data ingested or transmitted out of the app. The app usage can help to monitor and record users' actual behaviour based on the potential security flaws. An app based attack simulation approach can complement the survey-based results to compare the differences between self-reported behaviour (via survey) and actual behaviour (via usage).

**Threat II– Sources of Data Collection and Sample Size:** As per the study protocol (Phase 1, Figure 1), the data for this study has been collected via face-to-face meetings with the mHealth app users of two different mHealth provider (i.e., KFMC, HMG in Phase 3, Figure 2). The face-to-face sessions for data collection allowed us to notice and clarify any ambiguities faced by the participant of this study. The study's questionnaire could have been further validated using better pilot testing (e.g., semi-structured interviews, focus group meetings). We qualitatively analyzed the data using different statistical methods (Phase 4, Figure 1); however; the study results can be affected by the sample size. Also, the policy and regulatory measures regarding the privacy of health-critical data, consent of respondents, and capturing their feedback hampered us to collect more data. Moreover, all the end-users including, medics, health unit managers, clinical staff were based in the Saudi Arabia that limits geographical distribution of the participants (as indicated in Phase III of research method - Figure 1). Therefore, the presented results may slightly vary due to policies and regulations, practices to be followed, and employed end-users by mHealth providers. In order to minimise this threat, we plan to extend this study with collaboration from different mHealth providers and end-users at global level. A reader of this study should also consider that the results may also be influence by the policies and regulations, practices followed by mHealth services providers. In order to minimize this threat, we plan to extend this study by involving different mHealth providers and end-users at a global level scale.

## 7. CONCLUSIONS
mHealth apps have started to revolutionize the healthcare sector – empowering stakeholders such as governments, health units, medics, and patients – by offering context-sensitive and pervasive health services. Despite the offered benefits, mHealth apps are prone to security issues related to health-critical data and personal information of app users. We conducted an empirical study by using well established models like KAB and HAIS-Q to measure human aspects of mobile security in the domain of mHealth systems. Specifically, we collected, synthesized, analyzed, and documented responses from 101 end-users of two mHealth apps regarding their security-awareness towards app usage. The analysis of the demography data of the end-users' highlight that factors such as level of IT knowledge, age group, past experience with mHealth apps, mobile platforms they use, and educational backgrounds influence end users' security-awareness (i.e., users' knowledge, attitude, and behaviour ). The results suggest that end-users' security specific:
- Knowledge strongly influence their attitude towards security of mHealth apps. This means that users are aware of risks (e.g., stealing/tempering of health-critical data) and like to mitigate them (e.g., app support for data encryption).
- Knowledge does not significantly influence their behaviour. This means that users are aware of risks (e.g., private data shared with third parties for targeted ads) but are reluctant or unware of appropriate actions that mitigate risks (e.g., setting privacy preferences to restrict undesired data access).

## 8. ACKNOWLEDGMENTS

We thank the research centers at KFMC and HMG for approving our study. We would also like to thank all respondents for being a part of our study.